\documentstyle[aps,pra]{revtex}

\begin{document}
\title{Limitations of practical multi-photon decoherence-free states}
\author{Yong-Sheng Zhang\thanks{%
Electronic address: yshzhang@ustc.edu.cn}, Chuan-Feng Li, Yun-Feng Huang%
\thanks{%
Electronic address: hyf@ustc.edu.cn} and Guang-Can Guo\thanks{%
Electronic address: gcguo@ustc.edu.cn}}
\address{Laboratory of Quantum Information, University of Science and Technology of\\
China, Hefei 230026, P. R. of China\vspace{0.5in}}
\maketitle

\begin{abstract}
\baselineskip12pt It is shown in this paper that decoherence-free subspace
(DFS) of practical multi-photon polarization can not avoid the exponential
decoherence even in the same extra-environment if the photons are
frequency-anticorrelated. The reason lies in that the condition of
collective decoherence is not satisfied in this case. As an example, the
evolution of biphoton's decoherence-free state is given. Possible solution
for feasible multi-photon's DFS state is also given.

PACS number(s): 03.67.Hk, 42.50.-p, 03.67.Pp \medskip
\end{abstract}

\baselineskip12pt

\section{Introduction}

Quantum information processing provides secure communication\cite{QKD,QKD2}
and powerful quantum communication such as quantum teleportation\cite{Tel}.
The advantages of these tasks originate from the superposition of quantum
states\cite{Nielsen}. However, superposition is very fragile and easily
destroyed by the decoherence process due to unwanted coupling with the
environment\cite{Deco}. Several strategies have been devised to cope with
decoherence, each of them is appropriate for a specific type of coupling
with the environment\cite{codeco}. The first, Quantum error correcting code%
\cite{QECC}, relies on trying to detect errors using ancillary quantum bits
(qubits) and actively manipulating the interactions to correct these errors.
The second strategy employs dynamical decoupling\cite{Zeno,zeno2}, in which
rapid switching is used to average out the effects of a relatively slowly
decohering environment. The final approach attempts to embed the logical
qubits into a part of the overall Hilbert space that is inherently immune to
noise, a decoherence-free subspace (DFS)\cite{DFS,DFS2}. Beside the general
solutions mentioned above, another strategy particularly for long-distance
quantum communication has been also presented \cite{repeater}, i.e. quantum
repeater which combines entanglement purification and entanglement swapping
to build reliable communication channel.

Photons are obvious candidates for mediators of quantum information
particularly in quantum communication since they are fast, achieved easily,
and interact weakly with the environment. The degree of freedom used to
encode the information can be the polarization of the photon, its phase, or
some combination of both. In decoherence process of photons, polarization
effects are important source of problems, e.g. polarization-mode dispersion
in optical fibre \cite{PMD,PMD2}.

Quantum error correcting code is not suitable to overcome the photon's
decoherence, since it needs at least five qubits to encode a logic qubit and
correct only single-bit error. Though five-photon entangled state has been
realized\cite{5photon}, achieving a five-photon state, encoding and decoding
it definitely is a hard task in recent future. The dynamical decoupling
strategy is also not so easy to be realized as the similar reasons.
Entanglement purification in quantum repeater also needs many steps of
quantum operations. Linear optical quantum repeater, which has been
demonstrated in experiment \cite{Exprepeater}, however needs all entangled
photon pairs are produced from the common sources, and this is not practical
for long-distance communication. It seems that DFS is a good method to
conquer the decoherence since it only needs two qubits to encode a state and
four qubits to encode a logic qubit and it does not need so many quantum
operations between different qubits. In fact, some schemes based on DFS have
been demonstrated by photons \cite{Kwiat,Exp} or other systems \cite{Exp2}
and much more promising quantum information protocols based on photon's DFS
states have been proposed, such as robust quantum key distribution protocol 
\cite{DFSQKD}, communication without a shared reference frame \cite
{DFScomm,DFSother} etc.

Practical entangled photons which are mainly produced by spontaneous
parametric down-conversion (SPDC) are frequency-anticorrelated \cite{SPDC},
so they do not experience collective decoherence in communication medium
(e.g. optical fibre) which is the necessary condition of DFS. Hence, it is
shown in this paper that though frequency-anticorrelated multi-photon can be
encoded into a DFS and experience the same extra-environment, they can not
avoid the exponential decoherence. The reason lies in that they have
different intra-environment.

This paper is organized as follows: in Sec. II, we describe the single
photons decoherence in a practical way by the method of Master Equation\cite
{Noise}; in Sec. III, the decoherence of a frequency-anticorrelated
biphoton's DFS state is shown; in Sec. IV, we discuss the general cases of
decoherence of multi-photon's DFS states and in Sec. V we summarize the
conclusion.

\section{Single photon's decoherence}

A single-photon's initial pure state can be represented by a pure product
state of polarization and frequency\cite{Kwiat}, 
\begin{equation}
\left| \Psi \right\rangle =\left( a_1\left| H\right\rangle +a_2\left|
V\right\rangle \right) \otimes \int d\omega A\left( \omega \right) \left|
\omega \right\rangle  \eqnum{1}
\end{equation}
with basis $\left| H\right\rangle $ (horizontal polarization) and $\left|
V\right\rangle $ (vertical polarization) denoted by 
\[
\left| H\right\rangle =\left( 
\begin{array}{l}
1 \\ 
0
\end{array}
\right) ,\text{ }\left| V\right\rangle =\left( 
\begin{array}{l}
0 \\ 
1
\end{array}
\right) 
\]
respectively, and $A\left( \omega \right) $ is the complex amplitude
corresponding to $\omega $, normalized so that 
\[
\int d\omega \left| A\left( \omega \right) \right| ^2=1. 
\]
Here we assume that the frequency spectrum has a Gaussian form 
\[
A\left( \omega \right) =\pi ^{-\frac 14}\sqrt{\frac 1\delta }\exp \left[ -%
\frac{\left( \omega -\omega _0\right) }{2\delta ^2}^2\right] . 
\]
The stochastic coupling of photon's polarization with frequency via
anisotropic medium (e.g. non-ideal single-mode fibre) can be viewed as many
local birefringence processes of random direction \cite{PMD}. The
Hamiltonian that describes this coupling can be written as 
\begin{equation}
H_I=g\left( \omega \right) \left( \omega -\omega _0\right) \left[ \sigma
_x\sum_{{\bf k}_1}\left( a_{{\bf k}_1}+a_{{\bf k}_1}^{\dagger }\right)
+\sigma _y\sum_{{\bf k}_2}\left( a_{{\bf k}_2}+a_{{\bf k}_2}^{\dagger
}\right) +\sigma _z\sum_{{\bf k}_3}\left( a_{{\bf k}_3}+a_{{\bf k}%
_3}^{\dagger }\right) \right] .  \eqnum{2}
\end{equation}
Here $\sigma _{x,y,z}$ are Pauli operators, $g\left( \omega \right) $ is
assume to be a constant $g$ and the three bathes denoted by ${\bf k}_1,{\bf k%
}_2$ and ${\bf k}_3$ respectively have the same statistic properties. For a
Markovian reservoir\cite{Noise}, we can obtain the master equation of the
single photon's polarization as follows. 
\begin{equation}
\frac{\partial \rho }{\partial t}=\frac \gamma 2\left( \sigma _x\rho \sigma
_x+\sigma _y\rho \sigma _y+\sigma _z\rho \sigma _z-3\rho \right) .  \eqnum{3}
\end{equation}
Here $\gamma $ is a constant which is dependent on the medium. The solution
of this equation is 
\begin{equation}
\rho \left( t\right) =\rho \left( 0\right) e^{-2\gamma t}+\frac 12I\left(
1-e^{-2\gamma t}\right) ,  \eqnum{4}
\end{equation}
where $I$ is the $2\times 2$ identity matrix. The single photon's
polarization state will be decohered to the complete mixed state
exponentially.

\section{Decoherence of two frequency-anticorrelated photons' DFS state}

The two frequency-anticorrelated photons' DFS state can be described as 
\begin{equation}
\left| \Psi \right\rangle =\frac 1{\sqrt{2}}\left( \left| HV\right\rangle
-\left| VH\right\rangle \right) _{12}\otimes \int d\omega B\left( \omega
\right) \left| \omega _0+\omega \right\rangle _1\left| \omega _0-\omega
\right\rangle _2,  \eqnum{5}
\end{equation}
where $\omega _0$ is equal to half of centre frequency of the pump and we
assume the down-conversion photons have a Gaussian spectrum (in fact it
depends on the filter) 
\[
B\left( \omega \right) =\pi ^{-\frac 14}\sqrt{\frac 1\delta }\exp \left[ -%
\frac{\omega ^2}{2\delta ^2}\right] . 
\]
Here we neglect the width of the pump's spectrum, because it has no effect
on the interaction Hamiltonian of polarization and frequency. Particularly,
in the case of continuous pump such as Ar ion laser, the pump's spectrum
width is far narrower than the down-conversion photon's width. The
Hamiltonian that describes the coupling of two photons with the same medium
is 
\begin{equation}
H_I^{\prime }=g\sum_i\omega \left( \sigma _{i,1}-\sigma _{i,2}\right) \sum_{%
{\bf k}_i}\left( a_{{\bf k}_i}+a_{{\bf k}_i}^{\dagger }\right) ,  \eqnum{6}
\end{equation}
where $i=x,y,z$ and $\sigma _{i,1}$($\sigma _{i,2}$) are Pauli operators of
photon 1(2). For a Markovian reservoir, we can obtain the master equation of
the two photons' polarization as follows 
\begin{eqnarray}
\frac{\partial \rho }{\partial t} &=&\gamma \sum\nolimits_i[\left( \sigma
_{i,1}-\sigma _{i,2}\right) \rho \left( \sigma _{i,1}-\sigma _{i,2}\right) 
\eqnum{7} \\
&&-\left( \sigma _{i,1}-\sigma _{i,2}\right) \left( \sigma _{i,1}-\sigma
_{i,2}\right) \rho -\rho \left( \sigma _{i,1}-\sigma _{i,2}\right) \left(
\sigma _{i,1}-\sigma _{i,2}\right) ].  \nonumber
\end{eqnarray}
For an initial state of Eq. (5), the solution of the master equation is 
\begin{equation}
\rho \left( t\right) =\frac 14\left( 1-e^{-8\gamma t}\right) I_4+\rho \left(
0\right) e^{-8\gamma t},  \eqnum{8}
\end{equation}
where $I_4$ is the identity operator for a $4$-dimension system and $\rho
\left( 0\right) $ is the polarization state of Eq. (5). The stationary state
is $\rho =\frac 14I_4$, any deviation from it will vanish exponentially.
Details of the solution are shown in the Appendix. It can be seen that the
exponential decoherence is inevitable and there is none DFS state under the
Hamiltonian shown in Eq. (6).

\section{Decoherence of general DFS states}

The evolution of an $N$-photon's polarization state in the same medium can
be described by the following Hamiltonian 
\begin{equation}
H_I^{\prime \prime }=g\sum_{i=1}^3\sum_{j=1}^N\Delta _j\sigma _{i,j}\sum_{%
{\bf k}_i}\left( a_{{\bf k}_i}+a_{{\bf k}_i}^{\dagger }\right) ,  \eqnum{9}
\end{equation}
where $\Delta _j=\omega _j-\omega _0$ is the deviation from it's centre
frequency of the $j$th photon and $\sigma _{i,j}$ are Pauli operators of the 
$j$th photon. The master equation under Markovian approximation is 
\begin{equation}
\frac{\partial \rho }{\partial t}=\gamma ^{\prime }\sum_{i=1}^3\sum_{j,k=1}^N%
\overline{\Delta _j\Delta _k}\left( \sigma _{i,j}\rho \sigma _{i,k}-\sigma
_{i,j}\sigma _{i,k}\rho -\rho \sigma _{i,j}\sigma _{i,k}\right) .  \eqnum{10}
\end{equation}
It can be verified that the DFS state keeps invariant only when $\Delta _j=0$
holds for any $j.$ For example, in a four-qubit DFS the logic qubit is
denoted as 
\begin{eqnarray*}
\left| 0_L\right\rangle &=&\frac 12\left( \left| HV\right\rangle -\left|
VH\right\rangle \right) _{12}\left( \left| HV\right\rangle -\left|
VH\right\rangle \right) _{34}, \\
\left| 1_L\right\rangle &=&\frac 1{\sqrt{3}}\left( \left| HHVV\right\rangle
+\left| VVHH\right\rangle \right) _{1234} \\
&&-\frac 1{2\sqrt{3}}\left( \left| HV\right\rangle +\left| VH\right\rangle
\right) _{12}\left( \left| HV\right\rangle +\left| VH\right\rangle \right)
_{34}.
\end{eqnarray*}
There are $H_I^{\prime \prime }\left| 0_L\right\rangle \neq 0$ and $%
H_I^{\prime \prime }\left| 1_L\right\rangle \neq 0$ unless $\Delta _j=0$ for
any $j$.

\section{Discussion and summary}

The most important condition required by DFS is the collective decoherence,
i.e. all qubits must experience the same environment and have the same
coupling with the environment. In the coupling of photons with dispersive
medium, the environment includes medium and the intra environment, i.e. the
frequency freedom of photons. However, frequencies of two photons from SPDC
are anticorrelated which does not satisfy the condition of collective
decoherence.

There are two methods to solve this problem. The first is to let the
coupling coefficients of two photons are inverse everywhere which has been
used to demonstrate DFS in experiment\cite{Kwiat}, but it is not practical
in the stochastic medium for long-distance quantum communication. The second
method is to use frequency-correlated photon pairs. Since the conversation
of energy, producing frequency-correlated multi-photon states seems to be a
hard problem. However, only a part of the photons are frequency-correlated
also can let the conversation of energy hold. For example, we can produce
the four-photon state in SPDC process, but only use two frequency-correlated
photons of all photons. Though the efficiency decreases, we can also fulfil
many interesting tasks in long-distance quantum communication \cite
{DFSQKD,DFScomm,DFSother}. Another method for producing frequency-correlated
photons has peen proposed \cite{CorreSPDC}. It is essentially to produce a
narrow-spectrum but not true frequency-correlated photon pair, and it may be
useful in the situation which has not too high requirement in long-distance
quantum communication.

Note that we have neglected the spectrum width of the pump in the SPDC
process. But there is not any positive result if it is accounted, although
it has no effect on the coupling, it will cause dispersion independent on
the polarization.

The frequency distribution also exists in the coupling of other systems such
as atom in cavity, ion trap and solid qubit systems with the environment. It
is an open question whether it will affect the efficiency of the DFS in
these systems.

In conclusion, considering the photon's other freedom is involved in the
coupling with the environment, the photons may not decohered collectively
though all of them pass through the same communication medium i.e. all paths
of decoherence do not cancel each other. Therefore they can not avoid the
exponential decoherence and it is a drawback for long-distance quantum
communication. However, frequency-correlated photon pairs may be useful to
solve this problem.

\begin{center}
{\bf Acknowledgments}
\end{center}

We thank You-Zhen Gui and Fang-Wen Sun for helpful discussion. This work was
supported by the National Fundamental Research Program (2001CB309300), the
National Natural Science Foundation of China (No. 10304017, 10404027), the
Innovation Funds from Chinese Academy of Sciences.\medskip

\begin{center}
{\bf Appendix: Calculation of the decoherence of two-photon's DFS state}
\end{center}

The Eq. (7) can be written in this form 
\begin{eqnarray}
\frac 1\gamma \frac{\partial \rho _{11}}{\partial t} &=&-2\rho _{11}+\rho
_{22}+\rho _{33}-\rho _{23}-\rho _{32},  \eqnum{A-1} \\
\frac 1\gamma \frac{\partial \rho _{12}}{\partial t} &=&-3\rho _{12}+\rho
_{13}-\rho _{24}+\rho _{34},  \nonumber \\
\frac 1\gamma \frac{\partial \rho _{13}}{\partial t} &=&-3\rho _{13}+\rho
_{12}+\rho _{24}-\rho _{34},  \nonumber \\
\frac 1\gamma \frac{\partial \rho _{14}}{\partial t} &=&-2\rho _{14}, 
\nonumber \\
\frac 1\gamma \frac{\partial \rho _{22}}{\partial t} &=&-2\rho _{22}+\rho
_{11}+\rho _{44}+\rho _{23}+\rho _{32},  \nonumber \\
\frac 1\gamma \frac{\partial \rho _{23}}{\partial t} &=&-\rho _{11}+\rho
_{22}+\rho _{33}-6\rho _{23}-\rho _{44},  \nonumber \\
\frac 1\gamma \frac{\partial \rho _{24}}{\partial t} &=&-\rho _{12}+\rho
_{13}-3\rho _{24}+\rho _{34},  \nonumber \\
\frac 1\gamma \frac{\partial \rho _{33}}{\partial t} &=&-2\rho _{33}+\rho
_{11}+\rho _{44}+\rho _{23}+\rho _{32},  \nonumber \\
\frac 1\gamma \frac{\partial \rho _{34}}{\partial t} &=&\rho _{12}-\rho
_{13}+\rho _{24}-3\rho _{34},  \nonumber \\
\frac 1\gamma \frac{\partial \rho _{44}}{\partial t} &=&-2\rho _{44}+\rho
_{22}+\rho _{33}-\rho _{23}-\rho _{32}.  \nonumber
\end{eqnarray}
Note here 
\begin{equation}
\rho _{ij}=\rho _{ji}^{*}  \eqnum{A-2}
\end{equation}
for any $i,j$ and 
\begin{equation}
\sum_{i=1}^4\rho _{ii}=1.  \eqnum{A-3}
\end{equation}

To solve these equations, we can define the following variables 
\begin{eqnarray}
x_1 &=&\rho _{11}-\rho _{44},  \eqnum{A-4} \\
x_2 &=&\rho _{22}-\rho _{33},  \nonumber \\
x_3 &=&\rho _{11}-\rho _{22}-\rho _{33}+\rho _{44}+2\rho _{23}+2\rho _{32}, 
\nonumber \\
x_4 &=&\rho _{11}-\rho _{22}-\rho _{33}+\rho _{44}-\rho _{23}-\rho _{32}, 
\nonumber \\
x_5 &=&\rho _{23}-\rho _{32},  \nonumber \\
x_6 &=&\rho _{12}+\rho _{13},  \nonumber \\
x_7 &=&\rho _{24}+\rho _{34},  \nonumber \\
x_8 &=&\rho _{12}-\rho _{13}+\rho _{24}-\rho _{34},  \nonumber \\
x_9 &=&\rho _{12}-\rho _{13}-\rho _{24}+\rho _{34},  \nonumber \\
x_{10} &=&\rho _{14}.  \nonumber
\end{eqnarray}
There are 
\begin{eqnarray}
\stackrel{\cdot }{x_1} &=&-2\gamma x_1,\text{ }x_1=x_1\left( 0\right)
e^{-2\gamma t};  \eqnum{A-5} \\
\stackrel{\cdot }{x_2} &=&-2\gamma x_2,\text{ }x_2=x_2\left( 0\right)
e^{-2\gamma t};  \nonumber \\
\stackrel{\cdot }{x_3} &=&-8\gamma x_3,\text{ }x_3=x_3\left( 0\right)
e^{-8\gamma t};  \nonumber \\
\stackrel{\cdot }{x_4} &=&-2\gamma x_4,\text{ }x_4=x_4\left( 0\right)
e^{-2\gamma t};  \nonumber \\
\stackrel{\cdot }{x_5} &=&-6\gamma x_5,\text{ }x_5=x_5\left( 0\right)
e^{-6\gamma t};  \nonumber \\
\stackrel{\cdot }{x_6} &=&-2\gamma x_6,\text{ }x_6=x_6\left( 0\right)
e^{-2\gamma t};  \nonumber \\
\stackrel{\cdot }{x_7} &=&-2\gamma x_7,\text{ }x_7=x_7\left( 0\right)
e^{-2\gamma t};  \nonumber \\
\stackrel{\cdot }{x_8} &=&-6\gamma x_8,\text{ }x_8=x_8\left( 0\right)
e^{-6\gamma t};  \nonumber \\
\stackrel{\cdot }{x_9} &=&-2\gamma x_9,\text{ }x_9=x_9\left( 0\right)
e^{-2\gamma t};  \nonumber \\
\stackrel{\cdot }{x_{10}} &=&-2\gamma x_{10},\text{ }x_{10}=x_{10}\left(
0\right) e^{-2\gamma t}.  \nonumber
\end{eqnarray}
Combining equations from (A-2) to (A-5) we can obtain $\rho _{ij}\left(
t\right) $ for any $i,j$. The stationary state $\rho =\frac 14I_4$ can be
obtained if we let $x_{1,\cdots ,10}=0$, any deviation from it will vanish
exponentially.

\end{document}